\documentclass[sigplan,10pt,screen]{acmart}
\AtBeginDocument{%
  }

\usepackage{caption}
\usepackage{subfiles}
\usepackage{algorithm}
\usepackage{algpseudocode}
\usepackage{subfigure}
\usepackage{endnotes,microtype,graphicx,fancyvrb,multirow,diagbox}
\usepackage{xspace}
\usepackage{booktabs}
\usepackage[T1]{fontenc}
\usepackage{fancyhdr}
\usepackage{enumitem}
\usepackage[labelfont=bf,font=small,skip=5pt]{caption}
\usepackage{pifont}

\usepackage{fp}
\usepackage{siunitx}
\usepackage{tablefootnote}
\usepackage{tikz}
\usepackage{balance}
\usepackage{bbding}

\usepackage{soul}
\usepackage{zref-perpage}

\usepackage{framed}
\usepackage{mdframed}
\usepackage{multirow}
\usepackage{listings}

\renewcommand\footnotetextcopyrightpermission[1]{}

\newcommand{\sys}{\mbox{\textsc{Khaos}}\xspace}

\setcopyright{rightsretained}
\acmPrice{}
\acmDOI{10.1145/3579990.3580007}
\acmYear{2023}
\copyrightyear{2023}
\acmSubmissionID{cgo23main-p4-p}
\acmISBN{979-8-4007-0101-6/23/02}
\acmConference[CGO '23]{Proceedings of the 21st ACM/IEEE International Symposium on Code Generation and Optimization}{February 25 -- March 1, 2023}{Montréal, QC, Canada}
\acmBooktitle{Proceedings of the 21st ACM/IEEE International Symposium on Code Generation and Optimization (CGO '23), February 25 -- March 1, 2023, Montréal, QC, Canada}

\begin{document}
\author{Peihua Zhang}
\affiliation{%
  \institution{SKLP, Institute of Computing Technology, CAS}
  \city{}
  \country{}
}
\affiliation{
  \institution{University of Chinese Academy of Sciences}
  \city{Beijing}
  \country{China}
}
\email{zhangpeihua@ict.ac.cn}

\author{Chenggang Wu}

\affiliation{%
  \institution{SKLP, Institute of Computing Technology, CAS \& University of Chinese Academy of Sciences}
  \city{}
  \country{}
}

\affiliation{%
  \institution{Zhongguancun Laboratory}
  \city{Beijing}
  \country{China}
}
\email{wucg@ict.ac.cn}

\author{Mingfan Peng}
\affiliation{%
  \institution{SKLP, Institute of Computing Technology, CAS}
  \city{}
  \country{}
}
\affiliation{
  \institution{University of Chinese Academy of Sciences}
  \city{Beijing}
  \country{China}
}
\email{pengmingfan20g@ict.ac.cn}

\author{Kai Zeng}
\affiliation{%
  \institution{SKLP, Institute of Computing Technology, CAS}
  \city{}
  \country{}
}
\affiliation{
  \institution{University of Chinese Academy of Sciences}
  \city{Beijing}
  \country{China}
}
\email{zengkai@ict.ac.cn}

\author{Ding Yu}
\affiliation{%
  \institution{SKLP, Institute of Computing Technology, CAS}
  \city{}
  \country{}
}
\affiliation{
  \institution{University of Chinese Academy of Sciences}
  \city{Beijing}
  \country{China}
}
\email{yuding19s@ict.ac.cn}

\author{Yuanming Lai}
\authornote{Yuanming Lai is the corresponding author.}
\affiliation{%
  \institution{SKLP, Institute of Computing Technology, CAS}
  \city{}
  \country{}
}
\affiliation{
  \institution{University of Chinese Academy of Sciences}
  \city{Beijing}
  \country{China}
}
\email{laiyuanming@ict.ac.cn}

\author{Yan Kang}
\affiliation{%
  \institution{SKLP, Institute of Computing Technology, CAS}
  \city{}
  \country{}
}
\affiliation{
  \institution{University of Chinese Academy of Sciences}
  \city{Beijing}
  \country{China}
}
\email{kangyan@ict.ac.cn}

\author{Wei Wang}
\affiliation{
  \institution{SKLP, Institute of Computing Technology, CAS}
  \city{Beijing}
  \country{China}
}
\email{wangwei2021@ict.ac.cn}

\author{Zhe Wang}
\affiliation{%
  \institution{SKLP, Institute of Computing Technology, CAS}
  \city{}
  \country{}
}
\affiliation{%
  \institution{Zhongguancun Laboratory}
  \city{Beijing}
  \country{China}
}
\email{wangzhe12@ict.ac.cn}
\renewcommand{\shortauthors}{P Zhang, C Wu, M Peng, K Zeng, D Yu, Y Lai, Y Kang, W Wang, and Z Wang.}


\newenvironment{packeditemize}{
    \begin{list}{$\bullet$}{
        \setlength{\labelwidth}{8pt}
        \setlength{\itemsep}{0pt}
        \setlength{\leftmargin}{\labelwidth}
        \addtolength{\leftmargin}{\labelsep}
        \setlength{\parindent}{0pt}
        \setlength{\listparindent}{\parindent}
        \setlength{\parsep}{2pt}
        \setlength{\topsep}{3pt}}}{
    \end{list}
}

\pagestyle{fancy}

\newcommand{\bheading}[1]{{\vspace{4pt}\noindent{\textbf{#1}}}}

\newcommand{\ncircled}[1]{\textcircled{\raisebox{-.6pt}{#1}}}
\newcommand{\acircled}[1]{\textcircled{#1}}
\title{\sys: The Impact of Inter-procedural Code Obfuscation on Binary Diffing Techniques}

\begin{abstract}
Software obfuscation techniques can prevent binary diffing techniques from locating vulnerable code by obfuscating the third-party code, to achieve the purpose of protecting embedded device software. With the rapid development of binary diffing techniques, they can achieve more and more accurate function matching and identification by extracting the features within the function. This makes existing software obfuscation techniques, which mainly focus on the intra-procedural code obfuscation, no longer effective. 

In this paper, we propose a new inter-procedural code obfuscation mechanism \sys, which moves the code across functions to obfuscate the function by using compilation optimizations. Two obfuscation primitives are proposed to separate and aggregate the function, which are called fission and fusion respectively. A prototype of \sys is implemented based on the LLVM compiler and evaluated on a large number of real-world programs including SPEC CPU 2006 \& 2017, CoreUtils, JavaScript engines, etc. Experimental results show that \sys outperforms existing code obfuscations and can significantly reduce the accuracy rates of five state-of-the-art binary diffing techniques (less than 19\%) with lower runtime overhead (less than 7\%).

\end{abstract}

\maketitle

\vspace{-0.3em}
\section{Introduction}
\label{sec:intro}
\vspace{-0.3em}
Embedded devices have been widespread in many fields of modern life, such as wearables, traffic lights, and autonomous driving vision sensors and the total number are expected to reach 30 billion by 2025~\cite{iot_number}. In recent years, the number of vulnerabilities disclosed in embedded device software has been on the rise, and attacks targeting embedded devices have increased more than fivefold in the past four years~\cite{firmware_attack}. Once a vulnerability in an embedded device is exploited, it can lead to the collapse of the backbone network~\cite{antonakakis2017understanding}, while vulnerabilities in medical devices such as pacemakers are life-threatening~\cite{chen2020invisible,pacemaker}.

In addition to directly writing flawed code to introduce vulnerabilities, the reuse of vulnerable third-party code is another important reason for the widespread existence of vulnerabilities in embedded devices~\cite{nappa2015attack,jang2012redebug,cui2013firmware}. For example, Cui et al.~\cite{cui2013firmware} found that 80.4\% of LaserJet printers used third-party libraries with known vulnerabilities. However, vulnerabilities in these embedded devices cannot be patched in time due to the fragment issues --- similar code exists in multiple versions of various products due to the fast replacement of embedded devices~\cite{wei2016taming}. For example, the QualPwn vulnerability~\cite{qualcomm} in the Qualcomm's WiFi controller, which is equipped in millions of Android phones, took nearly 6 months from the vulnerability disclose to the patch released by the Qualcomm, and 
OEMs took longer to patch all devices across all versions.



Unfortunately, the above problem favors attackers in which they could detect existing vulnerabilities instead of exploring 0-day vulnerabilities laboriously. Since most embedded device software is not open source, attackers usually utilize the binary diffing techniques~\cite{cesare2013control,david2016statistical,feng2016scalable,david2018firmup,eschweiler2016discovre,hu2016cross,blazytko2017syntia,chandramohan2016bingo,feng2017extracting,hu2017binary,david2017similarity,xu2017neural,duan2020deepbindiff,BinDiff,huang2017binsequence,bourquin2013binslayer,luo2014semantics,ding2019asm2vec,zuo2018neural,gao2018vulseeker,wang2017memory,pewny2014leveraging,pewny2015cross,xu2020patch,zhao2020patchscope,xu2017spain,hu2013mutantx,ming2017binsim,jang2011bitshred,xu2017cryptographic,caliskan2015coding,zhan2021atvhunter,jin2012binary,xue2018accurate,wang2022enhancing,alrabaee2018fossil,hu2018binmatch,liu2018alphadiff,wang2022jtrans,haq2021survey} to locate the vulnerable code reused in the binary by comparing the binary with the third-party code. With the introduction of machine learning, binary differing techniques have made great progress in recent years. This greatly facilitates attackers locating existing vulnerabilities in binaries. For example, David et al.~\cite{david2018firmup} searched for common vulnerabilities in mobile devices, wearable devices, and medical devices, and were able to locate 373 existing vulnerabilities. 

Software obfuscation techniques~\cite{linn2003obfuscation,collberg2012distributed,banescu2016code,junod2015obfuscator,xu2018manufacturing,wang2020generating} can transform the program code to change the characteristics of the binary code even though the source code is the same. They could be used against binary diffing techniques, preventing attackers from locating existing vulnerabilities and thus protecting software. Recent researches have shown that software obfuscation techniques are no longer effective against the state-of-the-art binary diffing techniques~\cite{luo2014semantics,ming2017binsim,wang2017memory,ding2019asm2vec}. The main reason is that most software obfuscation techniques focus on the intra-procedural code obfuscation, which does not fundamentally change the semantics of functions, while binary diffing techniques can more and more accurately extract features within functions to obtain their semantics. 





Based on the above observations, we argue that inter-procedural code obfuscation should be emphasized due to their ability to change function semantics at the binary level. To this end, we propose an inter-procedural code obfuscation technique, \sys, which moves the code across functions and utilizes the compiler’s optimizations to transform (obfuscate) the code. The core idea of \sys is that \emph{once the code is restructured among functions, the generated binary code after compilation optimizations can be very different}. To achieve the inter-procedural code obfuscation, \sys changes function code across functions by separating a function into sub-functions and aggregating functions into one. 

It is non-trivial that transform arbitrary functions in \sys due to the challenges posed by performance, correctness, and obfuscation effect. For example, 1) To balance the obfuscation effect with the performance overhead, choosing which code blocks within a function (or functions) to be separated (or aggregated) is a problem; 2) Rebuilding all control flow and data flow among functions after transformations (especially
the indirect function calls handling in the fusion) is difficult; 3) Aggregating functions deeply without affecting the functionality of each function is a problem.

To address these challenges, two obfuscation primitives are proposed in \sys --- the \emph{fission} primitive and the \emph{fusion} primitive.
The fission is used to separate a function into sub-functions, and the fusion is used to aggregate several functions into one. The fission and the fusion are two complementary primitives, in which the fission tries to obfuscate the function by itself and the fusion tries to obfuscate the function by other functions. Furthermore, these two primitives can also be used together to improve the obfuscation effect, that is, the sub-functions separated by the fission can be aggregated with other functions again.

The fission partitions the code region to a sub-function on the control flow of the function with the dominator tree as the granularity, and also combines the static cold/hot code analysis technique to achieve lower performance overhead. Since the define-use relationships of variables are changed from within a function to cross functions, the fission needs to rebuild the data flow by passing parameters. To minimize the performance degradation caused by parameters passing, we also propose a data-flow reduction mechanism to reduce the number of parameters of the sub-functions. The control flow (including the exception control flow) is also rebuilt by inserting the function calls that call to sub-functions and encoding the return values in the sub-function.

The fusion selects two functions with compatible return values and no variadic parameters for the aggregation. Compatible means converting between different data types without losing precision. The parameter list of the post-aggregation function is merged from these two functions. To avoid the inefficient way of passing parameters through the stack, we propose a parameter list compression mechanism to reduce the number of the parameters. To rebuild the control flow completely, we propose a tagged pointer mechanism, which attaches control bits on function pointers to decide the executed code when the aggregated functions are called indirectly. We also propose a trampoline mechanism to handle the function calls across modules. To further improve the obfuscation effect, the deep fusion method is proposed to aggregate innocuous basic blocks, whose execution does not affect the global memory state, from different functions together within the aggregated function.

\sys was implemented based on the LLVM framework. The experimental evaluations were conducted on the Linux/X86\_64 platform by using SPEC CPU 2006 \& 2017 C/C++ programs, CoreUtils, and 5 common embedded device software containing vulnerabilities. 
Five state-of-the-art binary diffing tools~\cite{BinDiff,gao2018vulseeker,ding2019asm2vec,massarelli2019safe,duan2020deepbindiff} were used to evaluate the effectiveness of \sys. The results show that \sys is not only effective but also efficient: the effectiveness experiments show that the accuracy of these binary diffing was reduced to be less than 19\%, and the ranking of the vulnerable functions decreased significantly; the performance experiments show that \sys incurs less than 7\% overhead on average. In summary, our contributions are as follows:
\begin{packeditemize}
\item \textbf{A novel inter-procedural code obfuscation mechanism.} We point out that the inter-procedural code obfuscation is necessary against the binary diffing techniques, and propose a new obfuscation mechanism, \sys, which could obfuscate the code across functions.
\item \textbf{The fission and the fusion primitives.} We propose two obfuscation primitives in \sys to move the code across functions. The fission separates a function into multiple functions, and the fusion aggregates multiple functions into one.
\item \textbf{New insights from implementation and evaluation.} We
implement and evaluate a prototype of \sys, and the results show that it outperforms the existing obfuscators against the state-of-the-art binary diffing techniques. Our study suggests that binary diffing techniques should focus more on extracting the inter-procedural code features.
\end{packeditemize}
\section{Background and Motivation}\label{sec:bkg}
\subsection{Binary Diffing}\label{subsec:binary_diffing}
Binary diffing is a technique for visualizing and identifying differences between two binaries. It can quantitatively measure the differences between two given binaries and give matching result at predefined granularity (e.g., function). It has been widely used in software vulnerability search~\cite{pewny2014leveraging,pewny2015cross,gao2018vulseeker,david2018firmup,feng2016scalable,xu2017neural,chandramohan2016bingo,eschweiler2016discovre,zuo2018neural,david2014tracelet,zhan2021atvhunter,feng2017extracting,david2016statistical}, security patch analysis~\cite{xu2020patch,zhao2020patchscope,xu2017spain}, malware detection~\cite{cesare2013control,hu2013mutantx,ming2017binsim,jang2011bitshred,xu2017cryptographic}, code clone detection~\cite{luo2014semantics,caliskan2015coding,alrabaee2018fossil,hu2017binary,hu2018binmatch}, etc.

The workflow of binary diffing can be divided into two stages, i.e., the offline features extraction and the online code search. On the offline stage, tools extract features from binaries, while what features should be extracted is the focus of recent research; On the online stage, tools calculate the similarity of the given binaries by using extracted features. For example, BinDiff~\cite{BinDiff} extracts the number of basic blocks, control flow edges, and function calls within a function as the function's identity. Then, it combines the control flow graph matching algorithm to search for similar functions.

\subsection{Software Obfuscation}
Software obfuscation transforms the program without changing its functionality to make it hard to be analyzed. It can be used to hide vulnerabilities, protect intellectual property, etc. Actually, there is an arm race between software obfuscation and binary diffing. Software obfuscation does not want binary diffing techniques to match un-obfuscated with obfuscated code successfully, and vice versa. In recent decades, there are various techniques proposed in software obfuscation, and they can be classified into data obfuscation, static code rewriting, and dynamic code rewriting~\cite{schrittwieser2016protecting}.

Data obfuscation techniques~\cite{collberg1998breaking} transform the format of data to prevent it from direct matching. Since most binary diffing techniques utilize the features of the code, obfuscating data is less effective against binary diffing.

Various dynamic code rewriting approaches follow the concept of packing~\cite{nachenberg1997computer,roundy2013binary}, which hides code by encoding or encrypting it as data. But, the packing techniques are easy to be automatically unpacked~\cite{upx,themida} or be memory-dumped~\cite{egele2008survey,dinaburg2008ether,sharif2009automatic,ugarte2015sok}, which would lose the effect of obfuscation. Code virtualization is another popular obfuscation technique~\cite{king2006subvirt}. It translates code into specific interpret representations (IRs) instead of the native instructions and then uses an engine to interpret the IRs at runtime. This technique sacrifices much performance (10x slowdown at least~\cite{kuang2018enhance}) in exchange for a more powerful obfuscation. Therefore, the dynamic code rewriting technique is not suitable for fighting against binary diffing due to less effectiveness or too much overhead.

In contrast, static code rewriting is a promising technique against binary diffing. It modifies program code during obfuscation without further runtime modifications, which is similar to compiler optimization. Researchers have proposed many techniques for static code rewriting. For ease of introduction, we categorize them by obfuscation granularity:

\bheading{Instruction level:} Instruction substitution ~\cite{collberg2012distributed,junod2015obfuscator} replaces the original instruction with equivalent instruction(s), such as replacing an ``\texttt{add}'' instruction with two ``\texttt{sub}'' instructions. O-LLVM~\cite{junod2015obfuscator} designed 10 different substitution strategies for arithmetic and logical operations. To increase the complexity of conditional branch instructions, opaque predicate techniques ~\cite{collberg2012distributed,junod2015obfuscator,xu2018manufacturing,tiella2017automatic,ming2015loop} were proposed. They add permanent true or false (e.g., $x^2$ != -1) conditions that do not affect the original control flow, which are frequently used against analytical techniques such as symbolic execution.

\bheading{Basic block level:} Bogus control flow ~\cite{collberg2012distributed,junod2015obfuscator,ollivier2019kill} inserts dead code into the original control flow and often utilizes opaque predicates to prevent these codes from being optimized away and executed, thereby ensuring the original functionality of the program.

\bheading{Function level:} Control flow flattening ~\cite{collberg2012distributed,junod2015obfuscator} converts the control flow of the function into the ``\texttt{switch-case}'' form, which is hard to be analyzed, and maintains the original jump relationship by controlling the values of the cases. To prevent being degraded back to the original control flow, the ``\texttt{case}'' relationship is also obfuscated (encrypted).

\subsection{Motivation}
As binary diffing techniques continue to advance, many static code rewriting techniques (referred to as \emph{code obfuscation} in the rest of the paper) with the intra-procedural granularity (i.e., instruction, basic block, and function) are no longer effective~\cite{luo2014semantics,ming2017binsim,wang2017memory,ding2019asm2vec}. The main reason is that intra-procedure code obfuscations do not fundamentally change the semantics of each function, while most binary diffing techniques are increasingly capable of extracting features within functions to understand their semantics.



Therefore, we argue that \emph{inter-procedural code obfuscations should be emphasized due to their ability to change function semantics at the binary level which is the key to defeating binary diffing techniques}: 1) For the binary diffing works that only consider the intra-procedural information, the inter-procedural code obfuscation can fundamentally defeat them because the code structures along with the semantics are significantly changed; 2) For the binary diffing works that take inter-procedure information into account, the inter-procedural code obfuscation can also defeat them because the inter-procedural information extracted, such as the types of function calls, the numbers of function calls, and the call graph, are also significantly changed after the obfuscation.

Our thinking is also hinted by the literature published from both the offensive and defensive sides: 1) most of binary diffing works have discussed the issues of function inline~\cite{bourquin2013binslayer,luo2014semantics,feng2016scalable,chandramohan2016bingo,huang2017binsequence,ding2019asm2vec,xu2021interpretation,alrabaee2018fossil,david2017similarity,david2014tracelet,eschweiler2016discovre,feng2017extracting,hu2016cross,hu2017binary,hu2018binmatch}, and many of them~\cite{feng2016scalable,david2018firmup,huang2017binsequence,xu2021interpretation,alrabaee2018fossil,david2014tracelet,eschweiler2016discovre,feng2017extracting,hu2017binary,hu2018binmatch} admitted that it would affect the accuracy of diffing; 2) Ren et al.~\cite{ren2021unleashing} found the function inline could reduce the binary similarity by approximately 10\%.

\section{Our Solution: \sys}\label{sec:design}
\subsection{Overview}
\begin{figure*}[h]
  \centering
  \includegraphics[width=\linewidth]{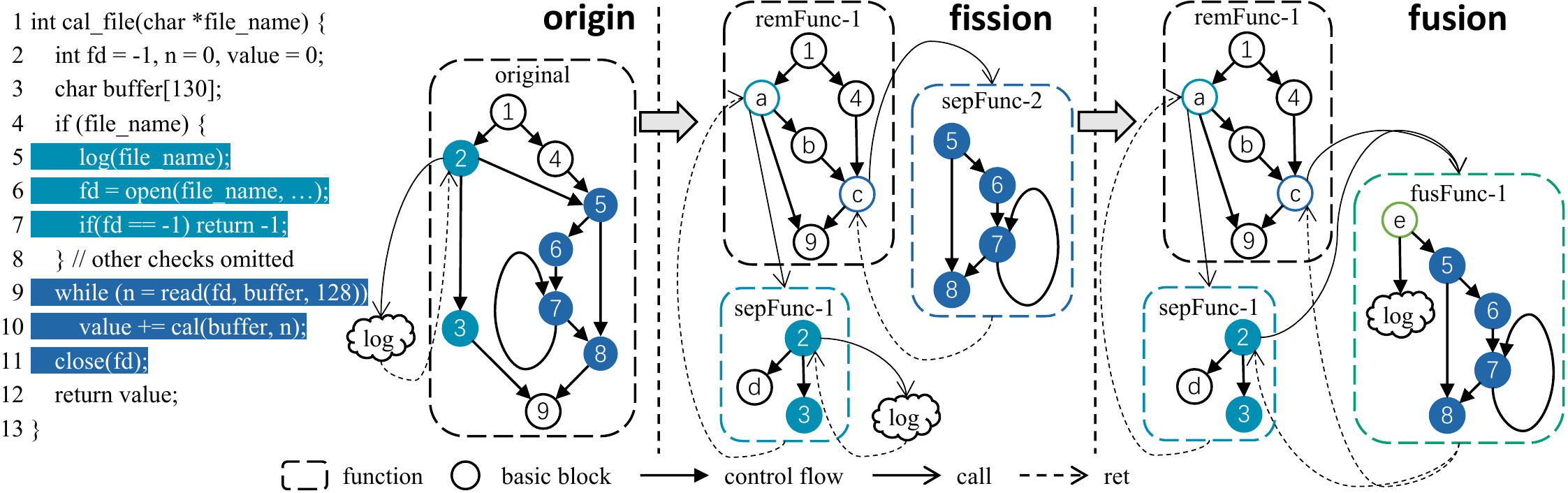}
  \caption{An example of obfuscating a function by using \sys.}
  \Description{.}
  \label{fig:overview}
\end{figure*}

To achieve the inter-procedural code obfuscation, \sys changes the amount of code within a function by moving code across functions firstly and then utilizes the compiler's optimizations to transform (obfuscate) the code. The idea behind it is that \emph{once the code is restructured among functions, the generated binary code after compilation optimizations (especially intra-procedural optimizations) can be very different}. 

In detail, we propose two obfuscation primitives --- the \emph{fission} primitive and the \emph{fusion} primitive. The fission primitive separates a function into multiple sub-functions thus making the function thinner. The fusion primitive aggregates functions into one thus making the function fatter. These two primitives can also be used together to make more in-depth changes to the function, that is, the separated sub-functions can be aggregated with other functions.

For the convenience of discussion, we denote a function before the transformation as an \textbf{oriFunc} (short for \emph{original function}), and denote the new function formed after the fusion as the \textbf{fusFunc} (short for \emph{fused function}). The new function formed by the separated code during the fission is denoted as the \textbf{sepFunc}
(short for \emph{separated function}), and the function formed by the remaining code  is denoted as the \textbf{remFunc} (short for \emph{remnant function}).

\autoref{fig:overview} gives an example about how the fission and the fusion are performed on a function named \texttt{cal\_file()}. The function is used to find the number of a special character in a given file. It first checks the file name and open the file (lines 4-7), then reads the content and counts the amount (lines 9-11). We can see that the fission separates two basic blocks (\ncircled{2}\ncircled{3}) to \texttt{sepFunc-1}, and four basic blocks (\ncircled{5}-\ncircled{8}) to \texttt{sepFunc-2}, respectively. To maintain the correctness, the fission inserts three trampoline basic blocks in the 
\texttt{remFunc-1} (\acircled{a}\acircled{b}\acircled{c}) to create the call relationship of two \emph{sepFunc}s. Basic block (\acircled{d}) is used to return different value of \texttt{sepFunc-1} (detailed in \autoref{chap:fission}). 
On top of the fission, the fusion aggregates the \texttt{log()} function and the \texttt{sepFunc-2} into a \texttt{fusFunc-1}. The entry basic block (\acircled{e}) will be inserted into the \texttt{fusFunc-1} to select the aggregated code blocks. 

Changing functions by recombining basic blocks from different functions is not trivial, and it still faces several challenges from performance, correctness, and obfuscation.
\begin{packeditemize}
\item \textbf{Challenge-1:} Choosing which basic blocks (or functions) to be separated (or aggregated) will seriously affect the performance overhead and obfuscation effect, and how to balance them well is difficult. For example, separating each basic block as a \emph{sepFunc} would favor the obfuscation, but brings unacceptable overhead.

\item \textbf{Challenge-2:} How to completely rebuild all control flow and data flow among functions after transformation (especially the fusion) is difficult. For example, 
once several functions participate in the fusion, we need to handle all pointers of the \emph{oriFunc}s so that it can correctly jump to the \emph{fusFunc} when de-referenced.

\item \textbf{Challenge-3:} 
Simply merging functions makes they become each other's junk code and has a limited obfuscation effect because the compiler will still optimize the code for different functions separately. Binding control flows and data flows belonging to different functions in the \emph{fusFunc} can prevent that but is also challenging to avoid changing the functionality of the function.
\end{packeditemize}

In the following subsections, we will detail the fission and the fusion design, and how we address the above challenges. 


\subsection{The Fission Primitive}
\label{chap:fission}
The fission first identifies the regions (each region is a basic block set) that need to be separated, then composes these regions into \emph{sepFunc}s, and finally rebuilds the control flow and the data flow among \emph{sepFunc}s and \emph{remFunc}.

\subsubsection{Partitioning Regions to Form sepFunc}
In general, a function's property is single entry and multiple exits. Hence, as long as a certain code region satisfies this property, it can be separated to become a new function. More precisely, as long as a code region is a dominator tree~\cite{allan1978domination} on the control flow graph, it can be extracted into a \emph{sepFunc}. The fission creates call relationship among \emph{sepFunc}s and \emph{remFunc} to ensure correctness. If the fission generates too many \emph{sepFunc}s, the newly created function calls in \emph{remFunc} will bring additional overhead (especially new function calls inside a loop). However, if the number or size of the \emph{sepFunc}s is small, the \emph{oriFunc} cannot be significantly changed. Therefore, designing a reasonable region identify algorithm is the key to reducing the overhead and improving the obfuscation effect.


\bheading{The core idea}. We abstract the code region partitioning problem as a graph cutting problem. The function's control flow graph can be regarded as a directed graph, and the edge weight represents the frequency of execution which indicates the cold/hot information. Partitioning the code region can be regarded as cutting the graph, where the weight of the cut edge is the cost of performance and the obfuscation effect is the number of the nodes in the sub-graph. 
\begin{algorithm}[!t]
\caption{The region identifying algorithm}\label{alg:cap}
\begin{algorithmic}[1]
    \Procedure{Identify}{$f$}\Comment{The function's representation}
        \State get dominator tree set $S$ of $f$
        \State $S \gets S\setminus {f}$\Comment{We won't separate the whole function}
        \While{$S$ is not empty}
            \State $target \gets$ null
            \For{dominator tree t in $S$}
                \State $effect \gets$ basic block count of $t$
                \State $cost \gets$ frequency of $t$'s head
                \If{$t$ is in loop}
                    \State $l \gets$ the innermost loop where $t$ is located
                    \State $cost \gets$ loop count of $l \times cost$
                \EndIf
                \State $value \gets effect \div cost$
                \If{$value>target$'s value}
                    \State $target \gets t$\Comment{Update the chosen tree}
                \EndIf
            \EndFor
            \If{$target\neq null$}
                \State delete trees from $S$ that intersect with $target$
            \EndIf
        \EndWhile
        \State \textbf{return}
    \EndProcedure
\end{algorithmic}
\label{alg:1}
\end{algorithm}

\bheading{The region identifying algorithm}. Based on the above idea, we design the region identifying algorithm (algorithm \autoref{alg:1}) on top of the directed weighted graph cut algorithm~\cite{stoer1997simple} to balance the performance overhead and the obfuscation effect. The algorithm takes function code as input and performs dominator tree analysis~\cite{lengauer1979fast} (line 2) at first. To avoid separating the whole function body into a \emph{sepFunc}, we remove the dominator tree of function itself (line 3) and identify the regions from the rest of the trees. To indicate the effect of the fission on obfuscation, we use the number of basic blocks in the tree to represent it (line 7). To indicate the effect of the fission on performance, we use the execution frequency of the root node of the dominator tree by using block frequency analysis~\cite{block_Frequency} (line 8) and the loop count (if the region is in a loop, the call to \emph{sepFunc} will increase) as the cost of the cut (lines 8-12). We iteratively select the most cost-effective (i.e., maximum the ratio of effect and cost) dominator tree to separate until the tree set is empty (lines 13-16).

\subsubsection{Data-flow Rebuild}
\begin{figure}[!t]
\centering
\includegraphics[width=\linewidth]{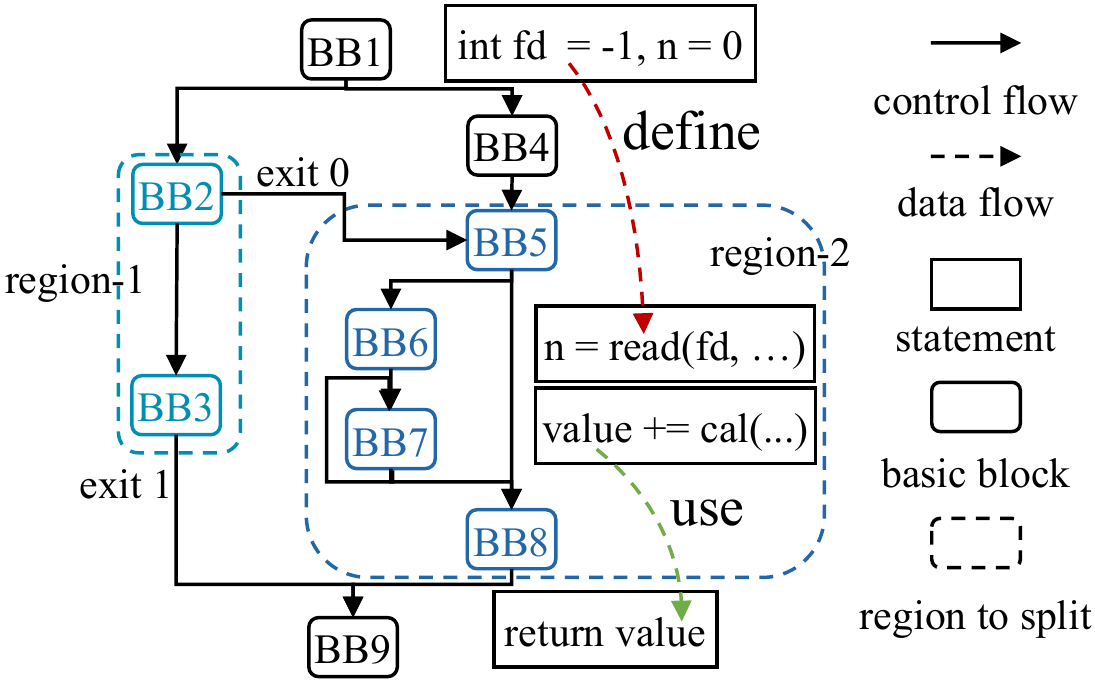}
\caption{The control-flow and data-flow graphs of cal\_file() in \autoref{fig:overview}}
\label{fig:fission_example}
\end{figure}

In addition to identifying regions as the function bodies of \emph{sepFunc}s, we also need to identify the inputs and the outputs of these regions to construct the parameters and return value of \emph{sepFunc}s. For each variable used in a region, it should be an input if its point is outside the region; Similarly, for each variable defined in a region, it should be an output if it has a use point outside the region. For example, as shown in \autoref{fig:fission_example}, the \texttt{fd} and \texttt{n} variables are inputs because the defined points are outside the region, and the \texttt{value} variable has a use point outside the region, so it is an output. For the variables whose define-use relationship are across regions, we use the function parameters to pass the pointer to them.
We don't pass a region's output variables by using the return value of \emph{sepFunc} because a region may have multiple output variables. 

\bheading{Data-flow reduction.} In general, the local variables of a function are defined at the entry basic block. Therefore, if an identified region needs to use local variables, these variables need to be passed into the \emph{sepFunc} through parameters.
In fact, if some local variables are only used by a \emph{sepFunc}, then these variables do not need to be passed into the \emph{sepFunc}, they can be defined directly in the \emph{sepFunc}. This can shorten the length of the \emph{sepFunc} parameter list, save unnecessary variable transmission, and further improve performance. To achieve this, we propose a lazy allocation strategy --- if a local variable is only used in the region, we will move the variable definition to the \emph{sepFunc}. For example, the \texttt{n} variable in \autoref{fig:fission_example} is initially defined in the \emph{oriFunc} but redefined and only used in the \texttt{region-2}, which becomes \texttt{sepFunc-2} function, so the definition point of the variable can be delayed in the \texttt{sepFunc-2} function.

\subsubsection{Control-flow Rebuild}
We extract the basic blocks of each identified region into a \emph{sepFunc}. The jump relationship between the regions in the \emph{oriFunc} is transformed into the function call-return relationship after fission. 
The creation of a function call is simple, we only need to insert the function call at the location of the entry basic block of the region before extraction and set the parameters that need to be passed into the \emph{sepFunc}.

The handling of function returns is relatively complex due to: If a region has multiple exits, the corresponding \emph{sepFunc} needs to encode this information into the return value, so that the \emph{remFunc} can use this information to select the corresponding code to execute. As \autoref{fig:fission_example} shows, for the two exits (\texttt{0} and \texttt{1}) in \texttt{region-1}, when \texttt{sepFunc-1} returns from \texttt{exit 0}, the control flow should go to \texttt{BB5}, and when returns from \texttt{exit 1}, it should go to \texttt{BB9}.

We use the return value of \emph{sepFunc} to indicate the \emph{remFunc} to determine the execution direction: We first number each exit of the \emph{sepFunc}, uses the number as its return value, and then insert a basic block at the call-site of this \emph{sepFunc} in the \emph{remFunc} (e.g., \acircled{a} in \autoref{fig:overview}) to transfer control flow based on the return value.

\begin{figure*}[!t]
  \centering
  \includegraphics[width=\linewidth]{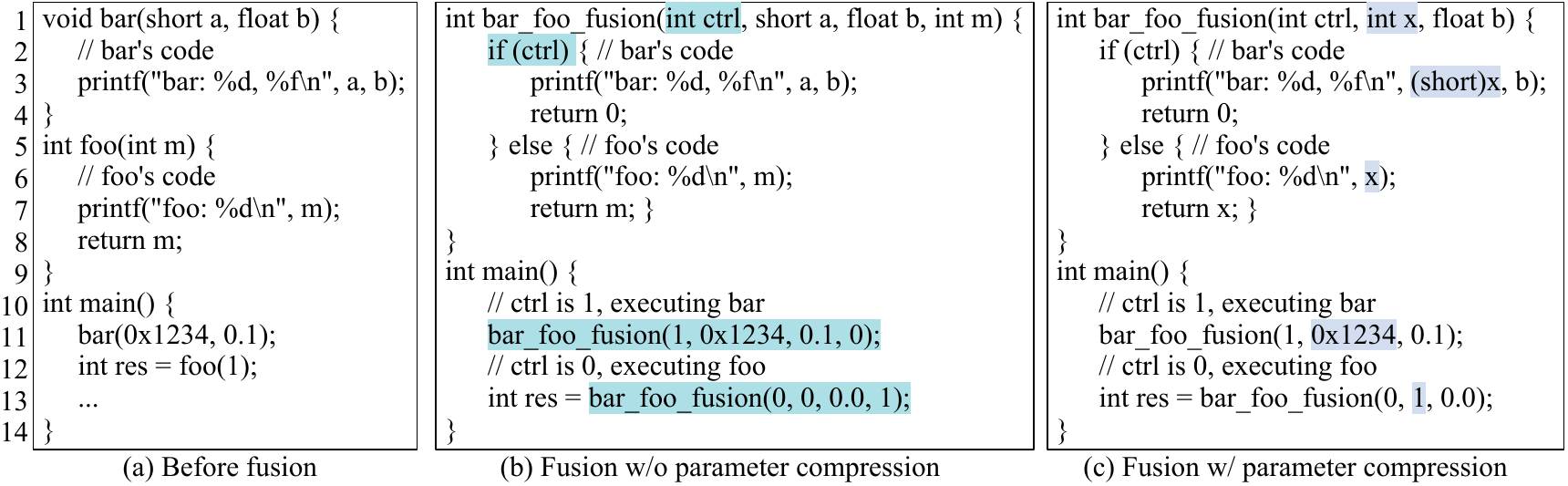}
  \caption{An example of performing the fusion on two functions.}
  \Description{.}
  \label{fig:fusion_example}
\end{figure*}

\subsubsection{Handling the Exception Control-flows} 
During program execution, there are some exception control flows that deviate from the usual function call and return, including the \texttt{setjmp/longjmp} and the C++ exception handling (EH in short). The fission requires special handles of them. 

\bheading{Handling the setjmp/longjmp.} Programmers could use the \texttt{setjmp()} in a function to record the current context into a \texttt{jmpbuf} structure. And then, they could use the \texttt{longjmp()} in any subroutines on the call chain of this function to go back to place the \texttt{jmpbuf} is pointing to, i.e., the call-site of the \texttt{setjmp()}. There is a requirement here that the \texttt{setjmp()} and the \texttt{longjmp()} using the same \texttt{jmpbuf} must be in the same call chain. Therefore, the call-site of the \texttt{setjmp()} cannot be separated into any \emph{sepFunc}, because the stack frame of the function that calling the \texttt{setjmp()} cannot be freed when the corresponding \texttt{longjmp()} is executed. Otherwise, the \texttt{longjmp()} will direct control flow to an unknown location.

\bheading{Handling the C++ exception.} The EH mechanism is a feature of the C++ that developers can capture exceptions in the \texttt{try} block by writing the \texttt{catch} statements. Since the fission moves part of the code into a \emph{sepFunc}, the \texttt{try-catch} pair may be broken, making EH information inconsistent. Simply skipping the exception-relevant function would reduce the obfuscation effect. Therefore, when identifying the code region, if it contains any code that may generate an exception, we will locate the corresponding \texttt{catch} code and divide it into the same region.

\subsection{The Fusion Primitive}
\label{chap:fusion}
The fusion selects functions to form \emph{fusFunc}, and rebuilds the control and the data flow to ensure the correctness.
In theory, the fusion can aggregate any number of functions. To balance the performance overhead and the obfuscation effect, we choose to aggregate two functions to form a \emph{fusFunc}. 

\subsubsection{Selecting Functions to Form fusFunc}
The fusion cannot arbitrarily select functions, it needs to select functions with compatible types of the return values. The definition of incompatibility is that if converting between two types loses precision, the two types are incompatible. For example, when the return value of one function is an integer and the other is a float, these two functions cannot be aggregated. 

In fact, there are other conditions that limit the selection of functions: 1) The variadic functions, e.g., the \texttt{printf(...)}; 2) Two functions with incompatible types of the return values; 3) Two functions that have a direct calling relationship. The first two constraints are designed for correctness, and the last is designed for performance to avoid generating a lot of recursive \emph{fusFunc}s. 
Functions that meet the above constraints will be randomly aggregated in pairs. 
\subsubsection{Data-flow Rebuild}
Once the two functions to be aggregated are determined, the function prototype of the corresponding \emph{fusFunc} can be determined immediately. For example, as shown in \autoref{fig:fusion_example} (a) and (b), the \texttt{bar()} and the \texttt{foo()} are aggregated into \texttt{int bar\_foo\_fusion()}. The \texttt{ctrl} parameter is used to select the function bodies aggregated from the \texttt{bar()} and the \texttt{foo()}. Determining the function prototype of \emph{fusFunc} is crucial to the rebuild of the data flow, which involves setting the parameter list and return value.

\bheading{Parameter list compression.} Simply merging the parameter lists of the two functions makes the parameter list of \emph{fusFunc} too long, which will degrade the performance of calling \emph{fusFunc}. This is because in the X86\_64 calling convention, the first six parameters are passed in registers, and the rest of the parameters are passed on the stack, which is an inefficient way. To achieve efficient parameter passing, we propose a parameter list compression mechanism --- if the types of the two parameters from the two functions are compatible, we compress them into one.
The reason why we can do this is that when a \emph{fusFunc} is called, only the parameter list of one of the functions participating in the aggregation
is used. For example, as \autoref{fig:fusion_example}(c) shows, both the \texttt{bar()} and the \texttt{foo()} have an integer parameter (\texttt{short a} and \texttt{int m}), we compress them into one integer parameter (\texttt{int x}).

If a parameter can not participate in the compression, it is copied into the parameter list of the \emph{fusFunc}. The number of parameters after the fusion will increase. In the worst case, it is the sum of the parameters of the two functions, which means none of the parameters can be compressed. To avoid using the stack to pass parameters as much as possible, we preferentially select functions with the total number of parameters less than six for the fusion.

\bheading{Return value determination.} Determining the return type of \emph{fusFunc} is relatively simple: 1) If the return type of one function is \texttt{void}, then the return type of the \emph{fusFunc} is the return type of another; 2) If the return types of the two functions are both not \texttt{void}, the compressed type is used as the return type of the \emph{fusFunc}, which is similar to the parameter list compression mechanism. 

\subsubsection{Control-flow Rebuild}
Once the \emph{fusFunc} is created, the two involved \emph{oriFunc}s need to be removed, and all call-sites to the \emph{oriFunc}s need to be replaced to call the \emph{fusFunc}. As mentioned before, a \texttt{ctrl} parameter will be added into the parameter list of the \emph{fusFunc} to select the code block aggregated from the \emph{oriFunc}s. The value of this parameter is \texttt{0} or \texttt{1}, which is set according to the original call-site of the \emph{oriFunc}. Since the \emph{fusFunc} parameter list includes the parameters of both \emph{oriFunc}s, we only need to pass the parameters required by the \emph{oriFunc} to the \emph{fusFunc} at the call-site of this \emph{oriFunc}. Unused parameters are set to be \texttt{0}.  

\begin{figure}[!t]
  \centering
  \includegraphics[width=\linewidth]{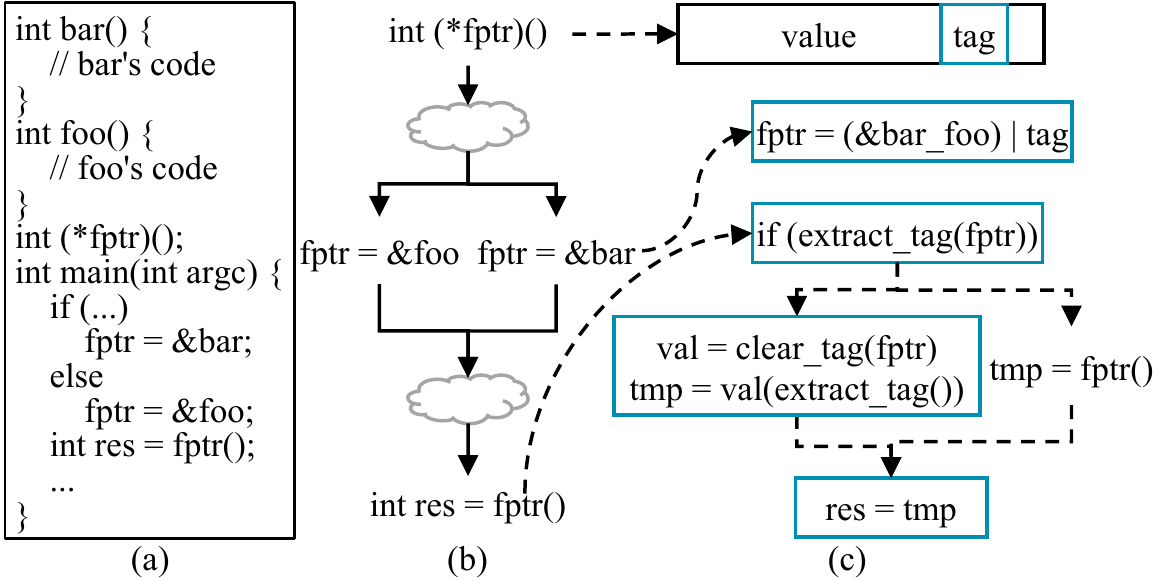}
  \caption{Function reference and indirect call processing.}
  \Description{.}
  \label{fig:indirect_call}
\end{figure}

\bheading{Handling Indirect function calls.} Indirect function calls are more difficult to handle than direct function calls because we do not know where the \emph{oriFunc} will be called. \autoref{fig:indirect_call} (a) shows an example that calls two functions by de-referencing the function pointer. The corresponding data flow is given in \autoref{fig:indirect_call} (b). When aggregating the \texttt{bar()} and the \texttt{foo()}, we need to change the function pointer points to the \emph{fusFunc} and then replace the function call to call this \emph{fusFunc}. But, we encounter a problem that we do not know what the value of the \texttt{ctrl} parameter should be set to. This is because at the compile time, we don't know whether the original function pointer \texttt{fptr} points to the \texttt{bar()} or the \texttt{foo()}.

To address the above problem, we propose a tagged pointer mechanism, which is similar to the low-fat pointer~\cite{kwon2013low}. The core idea is to encode the information (called \emph{tag}) of which \emph{oriFunc} pointed to by the original function pointer into the new function pointer, and when the new function pointer is de-referenced to make a call, the value of the \texttt{ctrl} parameter can be dynamically determined by parsing the new function pointer. 
In detail, when the operation of taking the address of the function participating in the aggregation occurs, we need to perform the encoding operation.
Since the \emph{tag} is encoded into the function pointer, it can be propagated along with the function pointer. When the function pointer is de-referenced to make a call, we will extract the \emph{tag} in the pointer as the and set the \texttt{ctrl} parameter according to the \emph{tag}.

The \emph{tag} requires two extra bits, where a bit indicates whether the pointer points to a \emph{fusFunc}, and the other bit records the value of the \texttt{ctrl} parameter. For example, as shown in \autoref{fig:indirect_call} (c), if the pointer \texttt{fptr} points to the \texttt{bar()}, the value of the \emph{tag} will be set to \texttt{11b}. When the pointer \texttt{fptr} is dereferenced to make a call, we insert code to first check whether the \emph{tag} is empty. If not, the code will extract the \texttt{ctrl} parameter and call the \emph{fusFunc}. Otherwise, no additional operations are required.

We choose the 2nd bit and the 3rd bit of function pointers to place the \emph{tag}. This is because the functions are usually 16-bytes aligned with the performance consideration, so the lowest 4 bits of the function pointer can be used (more reasons and considerations are detailed in \ref{chap:tag}).

\bheading{Handling function calls across modules.} There are two cases of cross-module function calls, one is the function pointer of a module is propagated to other modules, and the other is a module directly calls functions exported by other modules. If any case happens on a \emph{fusFunc}, we needs to process all involved modules to ensure the \emph{fusFunc} can be called correctly. But in some cases, we can not process all the modules (e.g., some libraries may have no source code).

To address this problem, we propose a trampoline mechanism so that all modules do not need to be processed. In detail, we transverse the data flow conservatively and identify all function pointers that may propagate outside the module. 
And then, we modify these function pointers to point to a piece of trampoline code instead of the \emph{fusFunc}. So that when the external module calls these function pointers, the control flow will transfer to the trampoline code first, and the trampoline code will help the function outside the module to reorganize the function parameters and call the \emph{fusFunc}. For the exported \emph{oriFunc}s, the method is similar to the replacing the \emph{oriFunc}'s function body with the trampoline code. 

\subsubsection{The Deep Fusion}
\label{chap:deep_fusion}
To further improve the obfuscation effect, we propose a deep fusion method to aggregate as many basic blocks as possible between the two parts of the code during the fusion process.

We have observed that \emph{some basic blocks can be executed many times without affecting the normal function}. The characteristic of these basic blocks is that their execution does not affect the global memory state, and they are called the \emph{innocuous basic block} in this paper. The concept is very similar to the \emph{reentrant function}~\cite{ralston2000encyclopedia} that it can be re-executed without affecting the functionality of the program. For innocuous basic blocks from different \emph{oriFunc}s, they can be aggregated together within the \emph{fusFunc}.
The innocuous analysis of each basic block is conservative. For example, 1) if a memory write operation in a basic block cannot be determined whether the modified data is local or global, then this basic block is not innocuous; 2) if there is a function call to an external function in a basic block, this basic block is not innocuous.

\begin{figure}[!t]
  \centering
  \includegraphics[width=\linewidth]{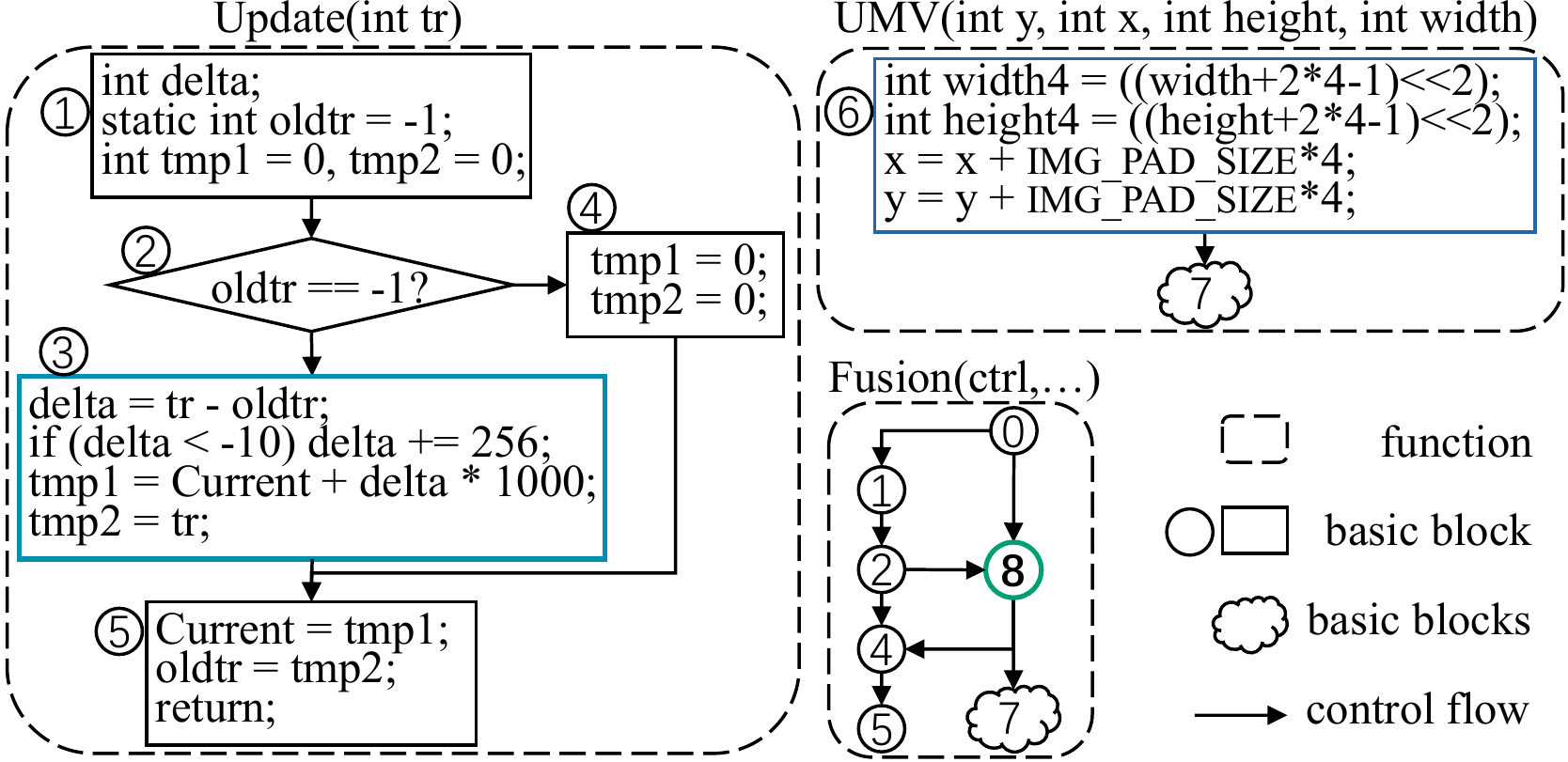}
  \caption{A real-world example of the deep fusion method.}
  \Description{464}
  \label{fig:deep_fusion}
\end{figure}

We give a simplified example of \texttt{464.h264ref} in SPEC CPU 2006 benchmark. As shown in \autoref{fig:deep_fusion}, the \texttt{Update()} and \texttt{UMV()} are aggregated into the \texttt{Fusion()}. The basic block (BB) \ncircled{3} of the \texttt{Update()} firstly redefines the local variable \texttt{delta}, and then loads the value of global variable \texttt{Current}, and writes two local variables \texttt{tmp1} and \texttt{tmp2} at last. Since these operations do not affect the global memory state, the BB \ncircled{3} is determined to be innocuous, and so as the BB \ncircled{6} of UMV(), thus we aggregate them into one --- the BB \ncircled{8}. 

This deep fusion method modifies the control flow graph and data flow graph of the \emph{fusFunc} at the same time, adding data dependency and control dependency so that the \emph{fusFunc} cannot be simply separated back to the two functions.

\subsection{Combining the Fission and the Fusion}
\label{chap:comb}
The fission and the fusion can be used together to further enhance the obfuscation effect. There are three combinations as follows:
\begin{packeditemize}
\item \textbf{FuFi.sep}: Only aggregating the \emph{sepFunc}s generated by the fission. In this case, the issue of handling indirect function calls no longer exists;
\item \textbf{FuFi.ori}: Only aggregating the \emph{oriFunc}s that are not processed by the fission, e.g., the functions with only one basic block. This combination could balance the obfuscation effect and the performance overhead well, and is suitable for software in most real-world scenarios;
\item \textbf{FuFi.all}: Aggregating the fission-generated \emph{sepFunc}s and the fission-unprocessed \emph{oriFunc}s uniformly and randomly. In this combination, the obfuscation effect is prioritized, followed by the performance overhead. It is suitable for programs that require a high obfuscation effect.
\end{packeditemize}

\vspace{-0.3em}
\section{Evaluation}\label{chap:evaluation}
\vspace{-0.3em}

\begin{figure*}[h]
  \centering
  \includegraphics[width=\linewidth]{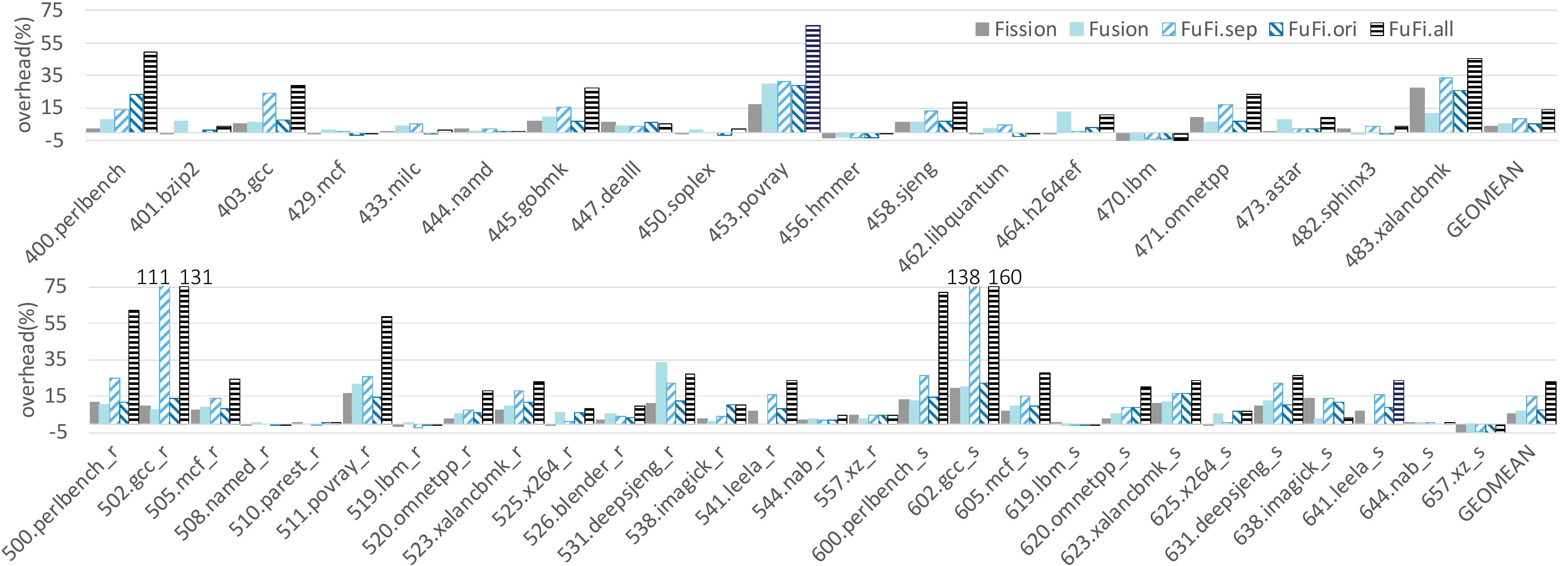}
  \caption{Runtime overhead of SPEC CPU 2006 (upper part) and 2017 (lower part) C/C++ programs.}
  \Description{}
  \label{fig:overhead}
\end{figure*}

We implemented \sys based on the LLVM-9.0.1. The fission and the fusion are implemented as the middle-end passes, and the fission pass is scheduled before the fusion pass. 
We run \sys on Ubuntu 20.04 (Kernel v5.4.0) that runs on an Intel(R) Xeon(R) Gold 5218 CPU with 128G memory. This section evaluates \sys in terms of effectiveness and performance, and answers the following questions:
\begin{packeditemize}
 \item (Q1) How is the performance of the obfuscated programs? 
 \item (Q2) How does \sys work against the state-of-the-art binary diffing techniques? 
 \item (Q3) How good is \sys at hiding real vulnerable code? 
 \end{packeditemize}

\bheading{Test Suites.} 
We used three test suites to evaluate \sys: 1) All C/C++ programs in SPEC CPU 2006/2017 benchmarks with the \emph{ref} input (denoted as the \textbf{T-I}); 2) All 108 programs in the CoreUtils 8.32 (denoted as the \textbf{T-II}); 3) Five commonly used programs in embedded devices with at least one vulnerability, including two popular IoT JavaScript engines (JerryScript and QuickJS), OpenSSL-1.1.1, BusyBox-1.33.1 and libcurl-7.34.0 (denoted as the \textbf{T-III}). The performance evaluation was performed on the \textbf{T-I} (Q1); The effectiveness against binary diffing techniques was evaluated on the \textbf{T-I} and the \textbf{T-II} (Q2); The ability to hide vulnerable code was evaluated on the \textbf{T-III} (Q3). Since software developers typically link programs into a single binary in embedded devices, we compiled and obfuscated these test suites in the same way under \texttt{O2} with the link-time optimization (LTO).

\bheading{Comparison targets.} 
To compare with existing obfuscator, we choose the popular compiler-level obfuscation tool O-LLVM~\cite{junod2015obfuscator} as our comparison target because it is open-sourced and compiler-based (same as \sys). O-LLVM~\cite{junod2015obfuscator} contains three obfuscation methods: instruction substitution~(\emph{Sub}), bogus control flow~(\emph{Bog}), and control flow flattening~(\emph{Fla}).  Literatures~\cite{ren2021unleashing,ding2019asm2vec,wang2017memory,banescu2016code} in software engineering, systems security, and programming languages fields all use it in their experiments. To ensure the consistency of the 
evaluation environment, we upgrade the LLVM version of O-LLVM~\cite{junod2015obfuscator} to 9.0.1, which is same as \sys. 
We also choose BinTuner~\cite{ren2021unleashing}, which is an iterative compiler tool that uses compiler options to transform the code to enlarge the difference of binaries, as another target to compare \sys with compiler's options.

\bheading{Confrontation targets.}
We use five state-of-the-art binary diffing techniques, i.e., Google BinDiff~\cite{BinDiff}, VulSeeker~\cite{gao2018vulseeker}, Asm2Vec~\cite{ding2019asm2vec}, SAFE~\cite{massarelli2019safe}, DeepBinDiff~\cite{duan2020deepbindiff}, to evaluate the effectiveness of \sys. Among them, Google BinDiff is an industry-standard binary diffing tool. Asm2Vec, SAFE, DeepBinDiff, and VulSeeker are the state-of-the-art methods for learning the semantic similarity in different granularity (e.g., function, basic block, control flow graph, call graph).

\subsection{Performance Overhead after Obfuscation}
We separately evaluated the performance overhead of the fission and the fusion, and the three combination modes introduced in \autoref{chap:comb} on the \textbf{T-I}. As shown in \autoref{fig:overhead}, the geometric performance overhead of the fission and the fusion are 5\% and 6\%, respectively. The reason why some cases (e.g., \texttt{456.hmmer}) have a negative performance overhead is that after the fission separates part of the code, the \emph{remFunc} can be further inlined to its callers, and the fusion improves the code locality of the aggregated functions. The results demonstrated that obfuscations compliant with the compiler optimizations can have good performance advantages.

Compared with the \emph{FuFi.ori}, the other two combinations have a higher overhead because the fission generates many \emph{sepFuncs}, aggregating them all incurs non-negligible performance overhead. For example, the \texttt{502.gcc\_r} contains many recursive functions, the \emph{sepFunc}s generated by these functions are aggregated to the \emph{fusFunc}s which are also the recursive functions. Since the 
stack frames of \emph{fusFunc}s are larger, they will bring much pressure to the stack. 

\begin{figure}[h]
  \centering
  \includegraphics[width=\linewidth]{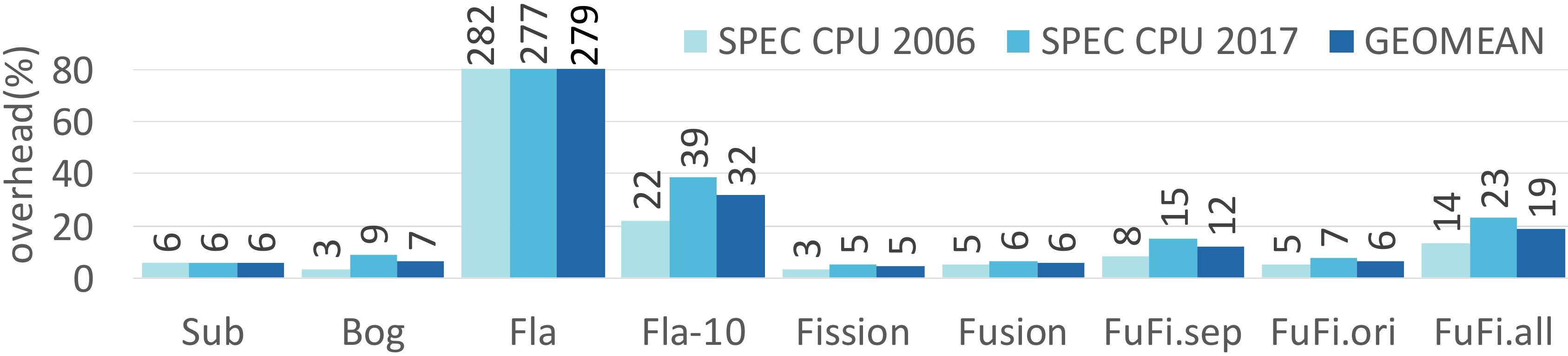}
  \caption{Runtime overhead of O-LLVM and \sys.}
  \Description{}
  \label{fig:overhead_compare}
\end{figure}

\bheading{Compared with O-LLVM\cite{junod2015obfuscator}.}
We compared the performance overhead of \sys with \emph{Sub}, \emph{Bog}, \emph{Fla}. As shown in \autoref{fig:overhead_compare}, \sys has comparable overhead with the \emph{Sub} and the \emph{Bog}. Due to the high overhead of \emph{Fla}, we reduce its obfuscation ratio to 10\% (\emph{Fla-10}), and others are all at 100\%.

\subsection{The Effectiveness against Binary Diffing}
\label{chap:reduce_accuracy}

Comparing binary diffing works is challenging due to their measurements of similarity are very different~\cite{ren2021unleashing}, such as graph edit distance or statistical significance. Simply comparing their similarity scores does not provide accurate information. For the commercial binary diffing tool BinDiff~\cite{BinDiff}, we normalized its similarity score to \texttt{[0, 1]}. For other tools open-sourced in academia, we normalized their results by computing the ratio of true matching function pairs that are also the top-ranked matching candidates (i.e. \emph{Precision@1}). 

\bheading{Paring success judgment method.}
Since \sys changes the number of functions, we relax the requirements for \emph{Precision@1}. For the fission, if the \emph{oriFunc} is paired with any \emph{sepFunc}s generated from it or the \emph{remFunc}, this pairing is 
recognized as successful. For the fusion, if the \emph{fusFunc} is paired with any function before the fusion, this pairing is recognized as successful. For the DeepBinDiff~\cite{duan2020deepbindiff}, since its result is basic block to basic block, the pairing is recognized as successful as long as their belonging functions are matched, even if the two basic blocks are not truly matched. It is worth noting that the above setting is looser than originally used in these tools but is more challenging for \sys.

\begin{table}
  \caption{Summarize of chosen diffing works.}
  \label{tab:summarize}
  \resizebox{1\columnwidth}{!}{
  \begin{tabular}{cccccc}
    \toprule
    & diffing & symbol & time & memory & call-graph\\
    & granularity & relying & consuming & consuming & lacking\\
    \midrule
    BinDiff~\cite{BinDiff} & function & Y & N & N & N\\
    VulSeeker~\cite{gao2018vulseeker} & function & N & Y & Y & Y\\
    Asm2Vec~\cite{ding2019asm2vec} & function & N & N & N & Y\\
    Safe~\cite{massarelli2019safe} & function & N & N & N & Y\\
    DeepBinDiff~\cite{duan2020deepbindiff} & basic block & N & Y & Y & N\\
  \bottomrule
\end{tabular}
}
\end{table}

\bheading{Test suite adjustment adaptability.} The characteristics of used binary diffing tools were summarized in \autoref{tab:summarize}. The column ``\emph{symbol relying}'' means the un-stripped binaries whether have side-effects or not, for example, BinDiff usually uses function names to reduce the searching space; The column ``\emph{time consuming}'' or the column ``\emph{memory consuming}'' means the diffing process takes a long time or requires a lot of memory; The column ``\emph{call-graph lacking}'' means whether using the call-graph as the feature or not.

\begin{figure}[!t]
  \centering
  \includegraphics[width=\linewidth]{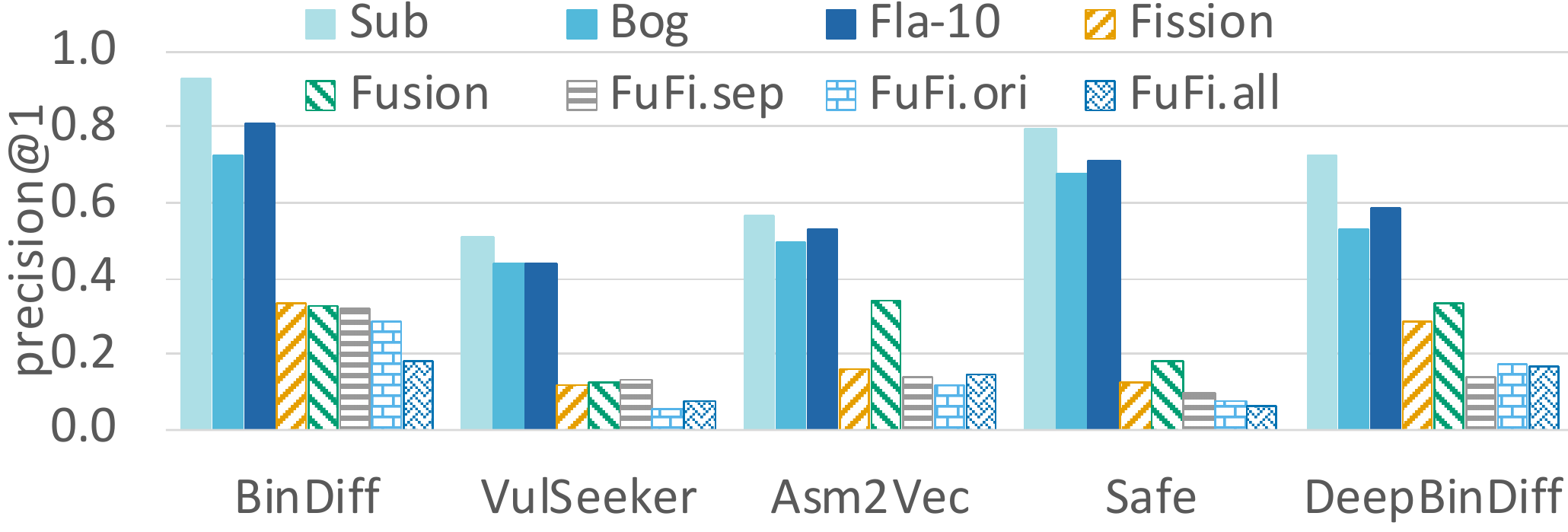}
  \caption{Precision@1 result of chosen binary diffing work.}
  \Description{Higher means lower adversarial effect.}
  \label{fig:precision_1}
\end{figure}

\begin{figure*}[h]
  \centering
  \includegraphics[width=\linewidth]{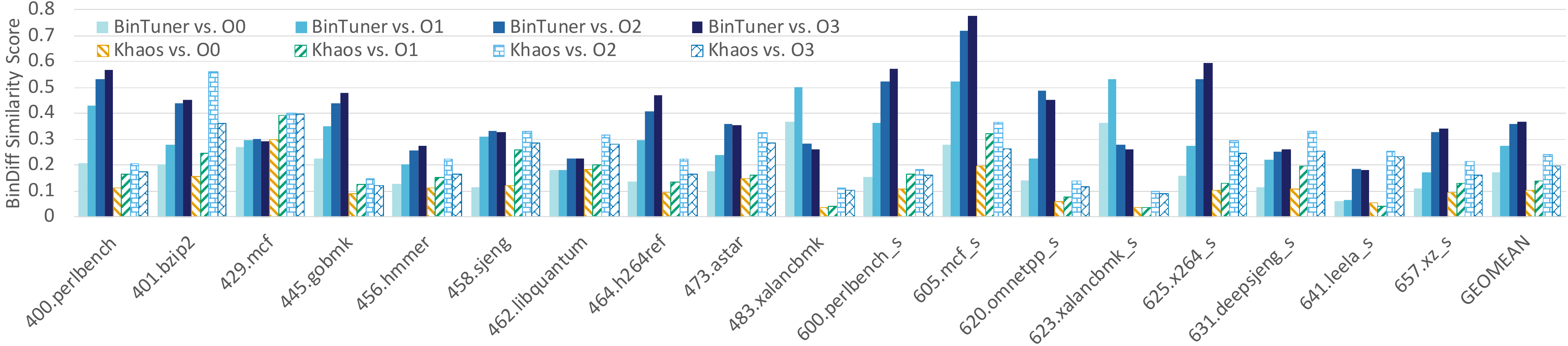}
  \caption{BinDiff similarity score of SPECint 2006, SPECspeed 2017 C/C++ programs.}
  \Description{}
  \label{fig:bintuner}
\end{figure*}

The test suites for VulSeeker~\cite{gao2018vulseeker} and DeepBinDiff~\cite{duan2020deepbindiff} need to be adjusted due to unable to run results. VulSeeker~\cite{gao2018vulseeker} takes more than 1 day to diff two large binaries and often gets killed due to memory limit. To speed up VulSeeker, we group the related functions into small groups (30 functions per group) to manually reduce the searching space, which is unfavorable to \sys because the smaller the group size, the easier to diff. DeepBinDiff~\cite{duan2020deepbindiff} requires too much memory (sometimes more than 10 TB) due to its representation of basic blocks. Since its diffing process is tightly coupled with binary size, we decide not to modify it and only use programs less than 40k lines. Even with the reduced test suite, it is still time consuming (e.g., over 1 week to diff binaries of \texttt{508.namd\_r}). It's worth mentioning that this is also unfavorable to \sys because it uses original functions to obfuscate each other, lacking material reduces the obfuscation effect. Other binary diffing tools still use the normal test suites.

\bheading{Results.}
We evaluated the accuracy of these tools by comparing obfuscated and un-obfuscated (un-stripped) binaries on the \textbf{T-I} and \textbf{T-II}. As \autoref{fig:precision_1} shows, higher accuracy means lower adversarial effect. Since BinDiff~\cite{BinDiff} takes the advantage of function names, its result is much higher than others. Although DeepBinDiff~\cite{duan2020deepbindiff} uses the basic block level instead of the function level as its granularity, the feature vector of the basic block still encodes the control flow graph and call graph, which have been changed by \sys, and that's why \sys can defeat it.
With comparable overhead, \sys can achieve a much better adversarial effect than O-LLVM~\cite{junod2015obfuscator}.

\bheading{Compared with compiler options.}
We follow the compare method of BinTuner\cite{ren2021unleashing} to calculate the similarity score of BinDiff\cite{BinDiff} under different compiler settings. For the BinTuner part, we set O0's binary code (same setting in the paper\cite{ren2021unleashing}) as the baseline during its iterative compilation. For the \sys part, we use binaries generated by \emph{FuFi.all}. As shown in \autoref{fig:bintuner}, \sys has a much lower similarity score in different compiler options. We also compared the overhead of programs generated by BinTuner with the baseline of \sys (O2 with LTO). The overhead is 30.35\%.

\subsection{The Ability of Hiding Vulnerable Code}
\label{chap:evaluation_3}
\begin{figure}[h]
  \centering
  \includegraphics[width=\linewidth]{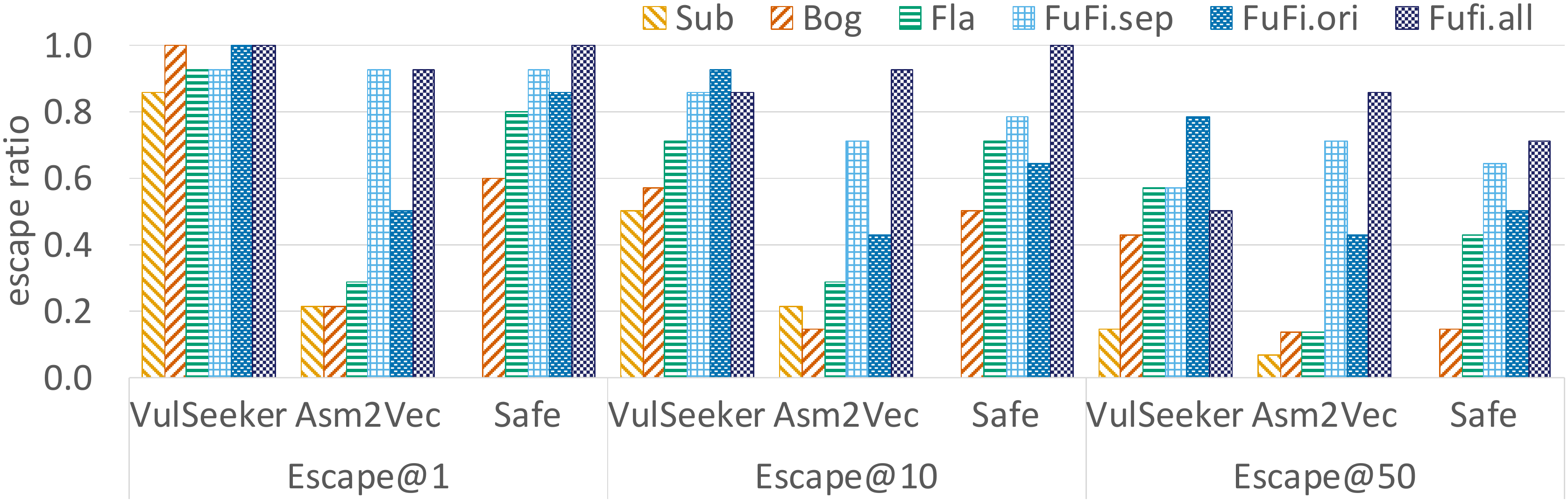}
  \caption{Escape ratio for top@1/10/50 of vulnerable functions. Higher means stronger hiding ability.}
  \Description{Higher means stronger adversarial effect.}
  \label{fig:escape}
\end{figure}

We use the \textbf{T-III} to further evaluate the ability of hiding real world vulnerable code. Each program contains at least one vulnerability (detailed in \ref{chap:cve}). In this experiment, we only used VulSeeker~\cite{gao2018vulseeker}, Asm2Vec~\cite{ding2019asm2vec}, and SAFE~\cite{massarelli2019safe} to calculate the \emph{escape@n ratio} (the rank of truly matched pair in the matched result) of vulnerable functions. The reason why BinDiff and DeepBinDiff were not used is that they only give top-1 matched result. We calculated \emph{escape@1/10/50 ratio} of vulnerable functions. For example, as shown in \autoref{fig:escape}, the  \emph{escape@50 ratio} of \emph{FuFi.all} on Asm2Vec is over 0.8, which means more than 80\% of vulnerable functions can not be found within top-50 ranked functions. Moreover, this time we set the obfuscation ratio of \emph{Fla} in O-LLVM to 100\%, which would bring unacceptable overhead in the real scenario.

\begin{table}[!t]
  \caption{Statistics of the fission and the fusion.}
  \label{tab:statistics}
  \resizebox{0.95\columnwidth}{!}{
  \begin{tabular}{cccc}
    \toprule
    & SPEC CPU 2006 & SPEC CPU 2017 & CoreUtils\\
    \midrule
    Fission Ratio & 116\% & 145\% & 152\%\\
    \#BB & 5.89 & 6.46 & 5.35\\
    RR & 34\% & 42\% & 44\%\\
    \midrule
    Fusion Ratio & 98\% & 97\% & 99\%\\
    \#RP & 1.27 & 1.43 & 1.47\\
    \#HBB & 1.08 & 1.02 & 1.89\\
  \bottomrule
\end{tabular}}
\end{table}

The \emph{escape ratio} could reflect the ability of hiding the vulnerable code with different obfuscations. With the same precision and binary diffing tool (e.g., \texttt{escape@50-} \texttt{Asm2Vec}), the \emph{FuFi.sep} and the \emph{FuFi.all} are better than the \emph{FuFi.ori}, and all of them are better than the \emph{Sub}, the \emph{Bog}, and the \emph{Fla} in O-LLVM. This ratio could also reflect the diffing ability of binary diffing tools. With the same precision and the settings of obfuscators, e.g., \texttt{escape@50-FuFi.all}, Asm2Vec is more accurate than Safe, and both of them outperform VulSeeker. 
The experimental results show that \sys can not only fight against binary diffing tools, but also reduce the pairing ranking of vulnerable functions significantly, achieving the purpose of hiding vulnerable code.

\subsection{The Statistics of \sys Internals}
\label{chap:evaluation_4}

\begin{figure*}[h]
  \centering
  \includegraphics[width=\linewidth]{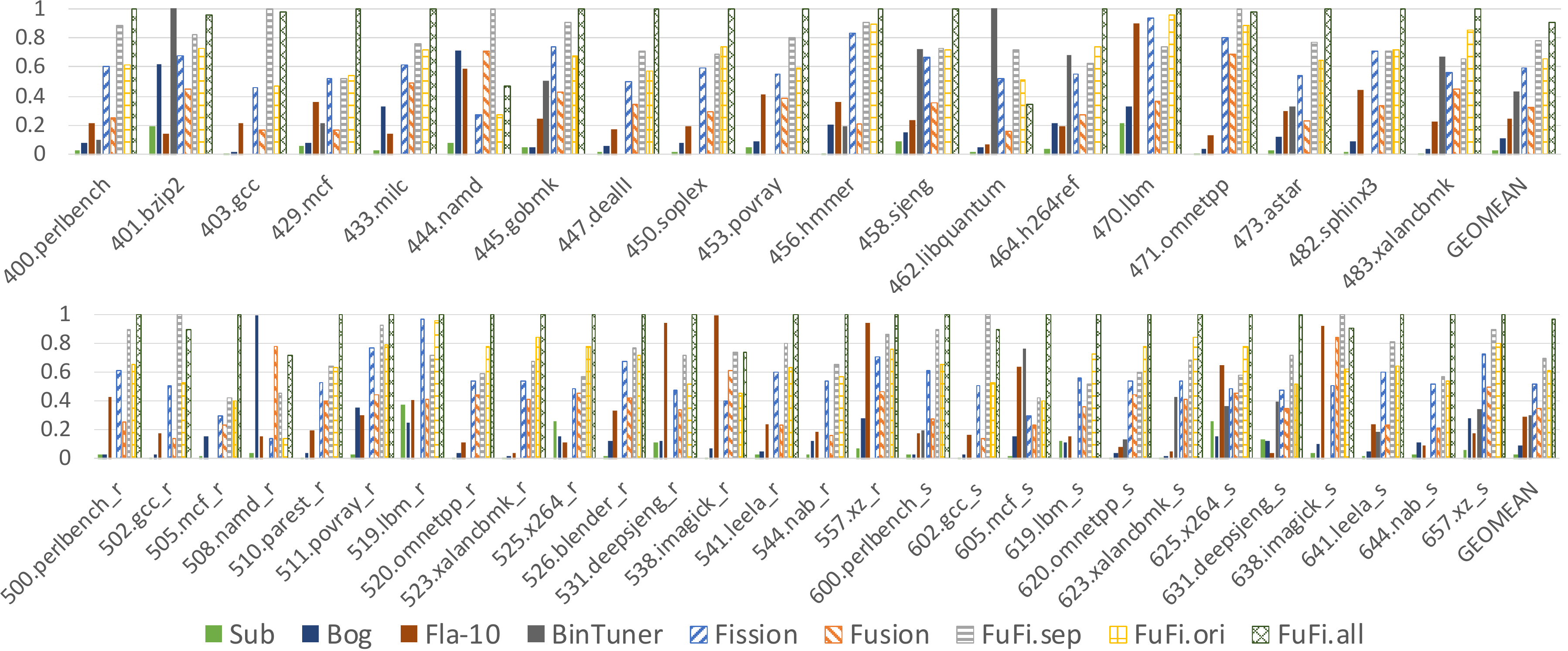}
  \caption{Opcode Histogram Distance (Normalized) for SPEC CPU 2006 \& 2017 C/C++ programs.}
  \Description{}
  \label{fig:histgram}
\end{figure*}

We collected some internal information on the \textbf{T-I} and \textbf{T-II} to demonstrate the effectiveness of \sys. We used the objdump tool to disassemble all the binaries and calculated the histogram of opcodes. After that, we calculated the vector distance between the origin and obfuscated binaries. Since different programs contain different amounts of codes, we used the max distance of all obfuscated programs as the baseline to normalize these distances. As shown in \autoref{fig:histgram}, the opcode distance of \emph{FuFi.all} is the largest, followed by \emph{FuFi.sep} and \emph{FuFi.ori}.

We also calculated the statistics of the fission and the fusion individually without the combination. For the fission, we counted the fission ratio (\#\emph{sepFunc}s / \#\emph{oriFunc}s), and the average number of basic blocks in \emph{sepFuncs} (\#BB), the reduced ratio of \emph{oriFuncs} after the fission (RR). For the fusion, we counted the fusion ratio (ratio of functions aggregated successfully), the reduced parameter number (\#RP) by parameter lists compression, and the number of innocuous basic blocks of each function (\#HBB). The statistical results are shown in \autoref{tab:statistics}.

These internal statistics proved that \sys can obfuscate the \emph{oriFunc}s with full force. For example, the \texttt{Fusion Ratio} is 97-99\%, which means almost all functions are aggregated. It also proved that both optimizations for runtime overhead (e.g., data-flow reduction) and obfuscation enhancement (e.g., innocuous analysis) have worked effectively.
\vspace{-0.6em} 
\section{Discussion and Future Work}
\vspace{-0.3em}
Aside from obfuscation techniques, we found that existing obfuscators have limitations on their implementation. In O-LLVM\cite{junod2015obfuscator}, \emph{Sub} can be optimized back under LLVM O3 option, which leads us to choose O2 as our baseline. \emph{Bog} and \emph{Fla} skip the exception-relevant functions. For Tigress\cite{collberg2012distributed}, we were unable to evaluate it in the same way as O-LLVM due to compilation errors.

The diffing process can be seen as a feature searching process. After we separate and aggregate these features, the searching difficulty increases and searching accuracy decreases. From our conclusion in table \ref{tab:summarize}, the lacking of call-graph consideration makes them unable to adopt inter-procedural obfuscation. We believe our study will raise awareness of inter-procedural obfuscation on binary diffing.

Smaller diffing granularity brings higher diffing costs. One way to reduce the cost is to use context information to narrow the searching space. Previous works pay much more attention to control flow rather than data flow. From the diffing perspective, data flow is harder to capture and encode. But from the obfuscation perspective, data flow is harder to change, too. Therefore, we predict the potential of data flow representation can be further tapped.

\section{Conclusion}
Binary diffing techniques can be used for 1-day/n-day vulnerability searching by attacker. In this paper, we propose an inter-procedural obfuscation technique \sys to protect software against the state-of-the-art binary diffing. We design two obfuscation primitives --- the fission and the fusion. Experimental results show that \sys is not only effective, but also efficient. We wish our study could not only help developers to protect their software, but also promote the development of binary diffing techniques in turn.

\begin{acks}
This research was supported by the National Natural Science Foundation of China (NSFC) under Grants 61902374, 62272442, U1736208, 61872386, and the Innovation Funding of ICT, CAS under Grant No.E161040.
\end{acks}
\appendix
\appendix

\section{Appendix}
\subsection{The tag bits choice.}
\label{chap:tag}

As mentioned in \autoref{chap:fusion}, the tagged pointer is used to select the code block aggregated from different \emph{oriFunc}s. On the X86\_64 architecture, only 48 bits of the virtual address are effective, so the upper 16 bits of the function pointer are unused and they can be used to place the \emph{tag} information for the fusion. But this approach is expensive when handling statically initialized pointers, such as global static function pointers and virtual function tables. For the position-independent executable, the values of these pointers need to be relocated to point to the actual function at load time. To attach the \emph{tag} information to these pointers, we need to add an initialization code to rewrite these pointers after the relocation which will make the program load slower.

To address the above problem, we choose to use the lowest bits of function pointers. This is because the addresses of functions are usually 16-bytes aligned with the performance consideration, so the lowest 4 bits of the function pointer can also be used to place the \emph{tag} information. Actually, the clang compiler has already used the least bit to identify whether a function pointer points to a virtual function or not, so currently, only the 3 bits are unused. Instead of rewriting statically initialized pointers after the relocation, we utilize the relocation mechanism directly by adding the \emph{tag}'s value to the \texttt{addend} field (which is used to add an offset when relocating) of the relocation item, so the \emph{tag} information can be attached to the pointer during the relocation. This method cannot be applied to support the upper bits \emph{tag} because it exceeds the range supported by the \texttt{addend} field, i.e., ($-2^{31}$, $+2^{31}$].

\subsection{CVE Detail}
\label{chap:cve}
\begin{table}
\caption{Vulnerable functions of Test Suite III}
\label{tab:cve}
\resizebox{\columnwidth}{!}{
\begin{tabular}{ c|c|c } 
\toprule
Program & Function & CVE\\
\hline
JerryScript & opfunc\_spread\_arguments & 2020-13991\\
\hline
QuickJS & compute\_stack\_size\_rec & 2020-22876\\
\hline
\multirow{2}{6em}{BusyBox1.33.1} & getvar\_s & 2021-42382\\ 
& handle\_special & 2021-42384\\ 
\hline
\multirow{2}{6em}{OpenSSL 1.1.1} & init\_sig\_algs & 2021-3449\\ 
& EC\_GROUP\_set\_generator & 2019-1547\\ 
\hline
\multirow{11}{6em}{libcurl 7.34.0} & suboption & 2021-22925,2021-22898\\
& init\_wc\_data & 2020-8285\\
& conn\_is\_conn & 2020-8231\\ 
& tftp\_connect & 2019-5482,2019-5436\\ 
& ftp\_state\_list & 2018-1000120\\
& alloc\_addbyter & 2016-8618\\
& Curl\_cookie\_getlist & 2016-8623\\
& \multirow{2}{8em}{ConnectionExists} & 2016-8616,2016-0755,\\ 
&  & 2014-0138,2015-3143\\
\hline
Total & 14 & 19 \\
\hline
\end{tabular}}
\end{table}

As discussed in \autoref{chap:evaluation_4}, we use the \textbf{Test Suite III} to further evaluate the ability of hiding real world vulnerable code. As shown in \autoref{tab:cve}, each program contains at least one vulnerability.
\balance
\bibliographystyle{ACM-Reference-Format}
\bibliography{ref}

\end{document}